# The gravitational energy for stationary space-time.

Roald Sosnovskiy
Technical University, 194021, St. Petersburg, Russia
E-mail:rosov2@yandex

**Abstract**.
It is prove, that the gravity field energy formulas obtained for static systems on the ground of local energy conservation law by test-particles fall, is suitable for stationary systems.

**1. Introduction.**

In [1] is proved the approach that allow obtain the formula of the static gravity field energy from the local energy conservation law. In [2] this approach was employ to non-symmetrical field.

This approach is essential connected with use of the coordinate system, motionless relatively to field source. This system essential differ from the all systems moving relatively to field source. It is necessary therefore to consider, is the obtained results suitable to others systems. In presented paper considered the gravitational energy in stationary space-time.

**2. The approach.**

The increase of static field energy cased by motion of the test-particles with rest masse $\delta m_0$ is equal [2]

$$dE_f = \frac{\delta m_0 c^2}{2\bar{g}_{00}} g_{00,i} dx^i \tag{1}$$

In [2] considered the static field of the asymmetric convex smooth infinitesimally thin material shell. It considered the space between a shell and some convex smooth external surface and the motion of $N_j$ discrete test-particles layers from the external surface to the shell. The motion is discrete; the number of steps is $N_q$. For every layer position, the calculations made for $N_k \cdot N_n$ points. Like that created is the net with $N_k \cdot N_n \cdot N_q$ points $P_{knq}$ and volume cells (k,n,q). Every cell have the edges $(\Delta z^1, \Delta z^1, \Delta z^1)_{k,n,q}$. The edges sours is the point $P_{knq}$. Through the every cell go $N_j$ test-particles layers rest mass

$$\delta m(k,n,q,j) = \delta\sigma(k,n,q,j) \cdot \Delta S(k,n,q,j) \tag{2}$$

Here $\Delta S(k,n,q,j)$ is the area of cell $(\Delta z^2, \Delta z^2)_{k,n,q}$ and $\delta\sigma(k,n,q,j)$ is the surface density in layer. The field energy change by intersection of the test-particle layer is equal

$$(dE_f)_{k,n,q,j} = \frac{c^2}{2}\left(\frac{\delta m_0 g_{00,\alpha} dx^\alpha}{g_{00}}\right)_{k,n,q,j}, \alpha = 1,2,3 \tag{3}$$

The field energy in cell (k,n,q) afterwards $N_j$ is equal

$$(dE_f)_{k,n,q} = \frac{c^2}{2}\sum_j (dE_f)_{k,n,q,j} \tag{4}$$

and the energy density

$$w_{k,n,q} = \frac{(dE_f)_{k,n,q}}{dV_{k,n,q}} \tag{5}$$

where $dV_{k,n,q}$ is the cell volume.

Total gravity energy in space is equal

$$E_f = \sum_{k,n,q}(dE_f)_{k,n,q} \tag{6}$$

The gravitating body mass is equal

$$m_0 = \sum_{n,q,j}\delta m_0(N_k,n,q,j) \tag{7}$$

For the coordinate system of the moving observer this quantity are called

$$d\hat{E}_f, \hat{E}_f, d\hat{V}, \hat{V}, \hat{w}, \hat{m} \tag{8}$$

**3. Criterions of the approach right.**

The energy (or mass) of the gravitational field in stationary space-time must depend on the velocity (*v*) the same as the field source mass. That signify, that the property of the energy not depend on its origin.

$$\hat{m} = \frac{m_0}{\sqrt{1 - v^2/c^2}} \tag{9}$$

$$d\hat{E}_f = \frac{dE}{\sqrt{1 - v^2/c^2}} \tag{10}$$

$$\hat{w}_f = \frac{w_f}{1 - v^2/c^2} \tag{11}$$

Also the ratio of the gravity field energy to the field sours mass must not depend on observer velocity

**4. Coordinates transformation.**

After the completion of the layer "j" motion it is necessary to calculate the metric $\bar{g}_{k,n,q,j}$ for all points $P_{k,n,q}$. By the asymmetrical field such calculation may be difficult, however it is possible, for example, by miens of the computer methods [3].

For cell $(\Delta z^1, \Delta z^1, \Delta z^1)_{k,n,q}$ the coordinate system $\bar{x}$ is transformed in the local system *x* with local diagonal metric

$$g_{00} = 1; g_{00,i} \neq 0; g_{\alpha\beta} = -\delta_\alpha^\beta; \alpha, \beta = 1,2,3$$

(here and further the indexes k,n,q,j my be omit, Latin indexes i,k…= 0,1,2,3,4, Greek indexes α,β …= 1,2,3)

This transformation hat the form

$$d\bar{x}^i = a_k^i dx^k, a_\alpha^0 = 0 \tag{12}$$

In this system the edges of cells is equal

$$\Delta x^\nu = c_\lambda^\nu \Delta z^\lambda, \Delta z^\lambda = h_\nu^\lambda \Delta x^\nu, c_\lambda^\nu h_\mu^\lambda = \delta_\mu^\nu \tag{13}$$

The observer system move relatively of the system *x* with velocity projections $v^\alpha$. In all points of the field the velocity module is same and equal *v*. As

$$\frac{\bar{v}^2}{c^2} = \frac{\bar{g}_{\alpha\beta} d\bar{x}_\beta d\bar{x}_\beta}{\bar{g}_{00} d\bar{x}_{00}^2} \tag{14}$$

*v* is invariant by transformation (12) and

$$v = \bar{v} \tag{15}$$

Coordinates $d\hat{x}^i$ in observer system is equal [4]

$$d\hat{x}^0 = \gamma dx^0 - \frac{\gamma}{c^2} v_\alpha dx^\alpha \tag{16}$$

$$d\hat{x}^\beta = dx^\beta - \gamma \frac{v^\beta}{c} dx^0 + (\gamma - 1) \frac{v_\alpha v^\beta}{v^2} dx^\alpha \tag{17}$$

$$\gamma^{-1} = \sqrt{1 - \frac{v^2}{c^2}} \tag{18}$$

In other form

$$d\hat{x} = Bdx \tag{19}$$

$$B = \begin{Vmatrix} \gamma & -\beta L^T \\ -\beta L & B_v \end{Vmatrix} \quad (20)$$

$$\beta = \frac{\gamma \cdot v}{c\sqrt{\gamma - 1}} \quad (21)$$

L is the vector with components

$$L_\alpha = \sqrt{\gamma - 1} \frac{v_\alpha}{v} \quad (22)$$

and

$$B_v = I - L \cdot L^T \quad (23)$$

The shortening of the cell edges and one projections can by calculated from (16),(17). The distance between the events on the segment $\Delta x^\alpha$ by time interval $d\hat{x}_0 = 0$ can by calculated from equalities

$$dx^0 = \frac{\gamma}{c^2} v_\alpha \Delta x^\alpha \quad (24)$$

$$\Delta \hat{x}^\beta = F_\alpha^\beta \Delta x^\alpha \quad (25)$$

Here $\Delta x^\alpha$ is the projection of any segment,

$$F_\beta^\alpha = \delta_\beta^\alpha + \lambda \frac{v_\alpha v^\beta}{c^2} \quad (26)$$

$$\lambda = (\gamma - 1)\frac{c^2}{v^2} - \gamma \quad (27)$$

Determinant F

$$\det F_\beta^\alpha = \sqrt{1 - v^2/c^2} \quad (28)$$

The cell volume in coordinates $x$ is equal [5]

$$dV = \det[\Delta_\beta x^\alpha] \quad (29)$$

where $\Delta_\beta x^\alpha$ is the projection of cells edge component on $x^\beta$. In coordinates $\hat{x}$ the cell volume is equal

$$d\hat{V} = \det[\Delta_\beta \hat{x}^\alpha] \quad (30)$$

From (25) and (30) can to obtain

$$d\hat{V} = \det(F \cdot dV) = \sqrt{1 - v^2/c^2} dV \quad (31)$$

$$\hat{V} = \sum d\hat{V} = \sqrt{1 - v^2/c^2} \quad (32)$$

**5. Check-up of the conformity to requirements (10),(11)**

5.1. For the cell P(k,n,q) on step j. From (3),(9),(25),(28) by transformation (the indexes k,n,q,j are omitted)

$$dx^\alpha = D_\beta^\alpha d\hat{x}^\beta, D = B_v^{-1} \quad (33)$$

and

$$dE_f = \frac{c^2}{2} \delta m_0 \frac{g_{00,\alpha} dx^\alpha}{g_{00}} = \frac{c^2}{2} \delta \hat{m} \sqrt{1 - \frac{v^2}{c^2}} \frac{\hat{g}_{00,\beta} d\hat{x}^\beta}{\hat{g}_{00}} = \sqrt{1 - \frac{v^2}{c^2}} d\hat{E}_f \quad (34)$$

From (31),(34)

$$d\hat{w} = \frac{d\hat{E}_f}{d\hat{V}} = \frac{dw}{1-\frac{v^2}{c^2}} \qquad (35)$$

The requirements (10),(11) are fulfilled.
5.2. For the cell P(k,n,q), step $N_j$ (the indexes k,n,q are omitted)
From (3),(9),(25),(28)

$$dE_f = \frac{c^2}{2}\sum_{j=1}^{N_j}(\delta m_0 \frac{g_{00,\beta}dx^\beta}{g_{00}})_j = \sqrt{1-\frac{v^2}{c^2}}d\hat{E}_f \qquad (36)$$

and from (31),(35)

$$d\hat{w} = \frac{d\hat{E}_f}{d\hat{V}} = \frac{dw}{1-\frac{v^2}{c^2}} \qquad (37)$$

The requirements (10),(11) are fulfilled.

5.3. For all field.
From (3),(7),(25),(28)

$$E_f = \frac{c^2}{2}\sum_{k,n,q,j}(\delta m_0 \frac{g_{00,\beta}dx^\beta}{g_{00}})_{k,n,q,j} = \sqrt{1-\frac{v^2}{c^2}}\hat{E}_f \qquad (38)$$

and

$$\hat{w} = \frac{\hat{E}_f}{\hat{V}} = \frac{w}{1-\frac{v^2}{c^2}} \qquad (39)$$

Thus, the requirements of the part 3 fulfilled and examined approach is right.

### 6. References
1. Sosnovskiy R. **gr-qc** 0507016
2. Sosnovskiy R. **gr-qc** 0607112
3. Lehner L. **gr-qc** 0106072
4. Möller C. *The theory of relativity.Moskow*,1975
5. Synge J.L.*Relativity: the general theory.* 1966